# On a Bipolar Model of Hyperbolic Geometry and its Relation to Hyperbolic Robertson-Walker Space

Harry I. Ringermacher[1]
General Electric Global Research Center , Schenectady, NY 12309

Lawrence R. Mead
Dept. of Physics and Astronomy, University of Southern Mississippi, Hattiesburg, MS 39406

Negatively curved, or hyperbolic, regions of space in an FRW universe are a realistic possibility. These regions might occur in voids where there is no dark matter with only dark energy present. Hyperbolic space is strange and various "models" of hyperbolic space have been introduced, each offering some enlightened view.  In the present work we develop a new bipolar model of hyperbolic geometry, closely related to an existing model – the band model – and show that it provides new insights toward an understanding of hyperbolic as well as elliptic Robertson-Walker space and the meaning of its isometries. In particular, we show that the circular geodesics of a hyperbolic Robertson-Walker space  can be referenced to two real centers – a Euclidean center and an offset hyperbolic center. *These are not the Euclidean center or poles of the bipolar coordinate system but rather refer to two distinct centers for circular orbits of particles in such systems*. Considering the physics of elliptic RW space is so well confirmed in the ΛCDM model with respect to Euclidean coordinates from a Euclidean center, it is likely that the hyperbolic center plays a physical role in regions of hyperbolic space.

I.  INRODUCTION

   Bipolar coordinates are an unusual system, typically not even considered in geometry (one professional geometer asked "What are those? " in one communication).  However, in physical applications, this system is sometimes advantageous, for example as a natural coordinate choice for a variety of electromagnetic applications of Laplace's equation[1]. and, as we shall show below, for negatively curved Robertson-Walker (RW) space in Cosmology.  Negatively curved RW space will arise in Cosmology in regions of voids where dark energy but no dark matter is present[2]. At a distance, from the point of view of the Hubble flow and comoving or "frozen" coordinates, only the expansion of 3-space is affected. Locally, however, one must revert to "thawed" coordinates.  One must therefore face the possibility of addressing particle dynamics in such regions. Particle geodesics behave oddly in hyperbolic space and various "models" of hyperbolic space have been introduced, each offering some enlightened view.  Below we first describe relevant features of three existing models; the half-plane, disk, and more recent band model. Then we develop a new bipolar model of hyperbolic geometry, closely related to the band model, and show that it provides new insights toward an understanding of hyperbolic as well as elliptic Robertson-Walker space and the meaning of its isometries.

---

[1]Communications to ringerha@ge.com

## II. CONFORMAL MODELS OF HYPERBOLIC GEOMETRY

There are three well-known conformal models of hyperbolic geometry[3]: the Poincaré half-plane, the Poincaré disk and the Minkowski model. The three models are isomorphic and each displays different insights of hyperbolic space under varying boundary constraints. We shall focus on the first two. For example, consider the Poincaré half-plane. This hyperbolic 2-space has the metric,

$$ds^2 = \lambda^2 \left( \frac{dx^2 + dy^2}{y^2} \right), \qquad (1)$$

and represents a plane of constant negative curvature, $-1/\lambda^2$, "the hyperbolic plane", described by Cartesian coordinates (x,y), excluding the x-axis. Geodesics are semi-circles centered on the x-axis and perpendicular to it, and lines perpendicular to it (Fig. 1).

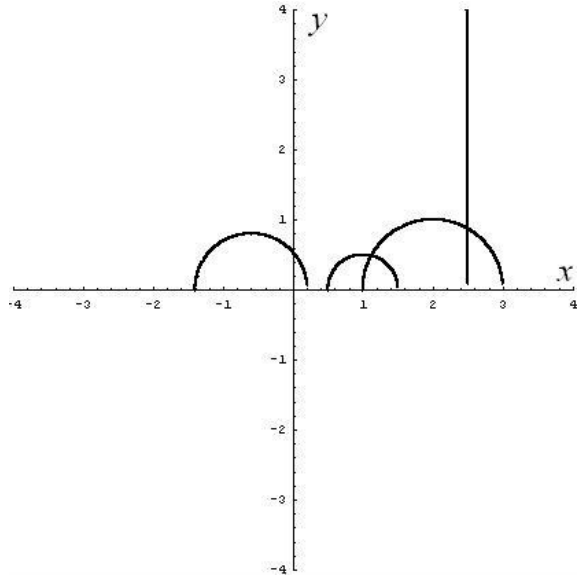

**Fig.1**. Geodesics of the Poincaré Half-Plane Model of the hyperbolic plane

The half-plane model has made a recent appearance in the Randall-Sundstrum $AdS_5$ theory putatively describing the "leakage" of gravity into the 5$^{th}$ dimension[4,5], which, in that case is a hyperbolic dimension replacing any physical dimension, y as above, or x, y, z, t in its generalization[6]. Recent work by Hubbard[7], describes a 4$^{th}$ conformal model of hyperbolic geometry, the band model.

The metric for this model is:

$$ds^2 = \frac{d\eta^2 + d\psi^2}{(\cos\psi)^2} \tag{2}$$

where $(\eta,\psi)$ are Cartesian coordinates. The geodesic equations are:

$$\begin{aligned}\eta'' + 2\eta'\psi'\tan\psi &= 0 \\ \psi'' - (\eta'^2 - \psi'^2)\tan\psi &= 0\end{aligned} \tag{3}$$

These equations yield two geodesics:

$$\eta = constant \quad \text{(vertical lines in Fig.2 – not plotted)} \tag{4}$$

$$\eta = \eta_0 \pm \left(\frac{1}{K}\right)\tanh^{-1}\left(\frac{-\sin\psi}{\sqrt{1-K^2(\cos\psi)^2}}\right) - \log\left(\sqrt{2}\left(-K\sin\psi + \sqrt{1-K^2(\cos\psi)^2}\right)\right) \tag{5}$$

$K$ and $\eta_0$ are constants. The Cartesian representation, $(\eta,\psi)$, of the complicated geodesics, Eq. (5), is plotted in Fig. 2. These are the geodesics desribed in Hubbard's work. The most useful property of this model is that the x-axis is included and is Euclidean. It is called the band model because the space of its complex representation is the band, $\pi/2 > \psi > -\pi/2$.

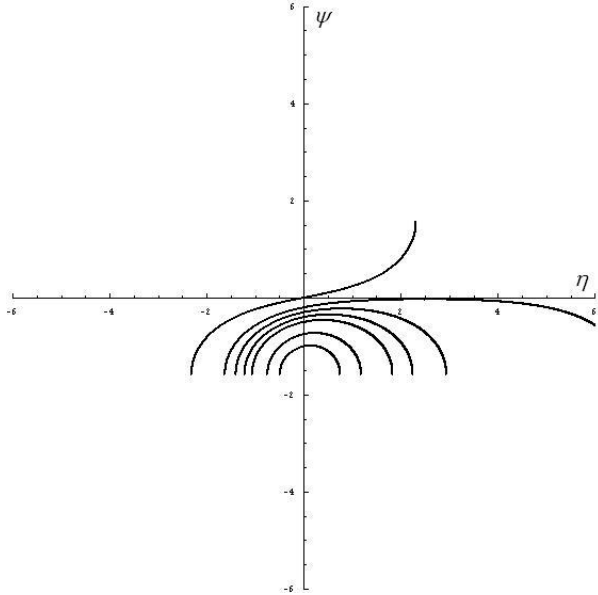

**Fig.2**. Geodesics of the Hubbard Band-Model of the hyperbolic plane

## III. THE BIPOLAR MODEL OF HYPERBOLIC GEOMETRY

We shall now prove the following proposition:

**Proposition 1**: The band model is the Euclidean bipolar representation of the Cartesian half-plane model.

**Proof:** Consider the half-plane metric in Cartesian coordinates $(x, y)$ of Eq.(1). Under the Euclidean coordinate transformation to a bipolar system $(x, y) \to (\eta, \psi)$,

$$x = \frac{h \sinh \eta}{\cosh \eta + \sin \psi}, \qquad y = \frac{h \cos \psi}{\cosh \eta + \sin \psi} \qquad (5)$$

for $h = 1$, $\lambda \equiv 1$, the metric (1) transforms to:

$$ds^2 = \frac{d\eta^2 + d\psi^2}{(\cos \psi)^2}. \qquad (6)$$

This is the band model metric (Eq.(2)) when the pair $(\eta, \psi)$ is interpreted as Cartesian and the proof is complete. We have shown that the band model is simply a different Euclidean view of the half-plane model. Rather than view $(\eta, \psi)$ as Cartesian, it is natural here to take them as the Euclidean bipolar coordinates defined with respect to Cartesian coordinates from Eq.(5). This is an entirely new representation of the half-plane or band model which we shall call the *bipolar model* of hyperbolic geometry. For convenience, we rotate the metric, Eq.(6), for a bipolar representation on the x-axis.

$$ds^2 = \frac{d\eta^2 + d\psi^2}{(\sin \psi)^2} \qquad (7)$$

The geodesics of the bipolar model are semi-circles about the poles at $h = \pm 1$ on the x-axis. The polar equations of these circles with respect to $(x, y) = (0, 0)$ are given by:

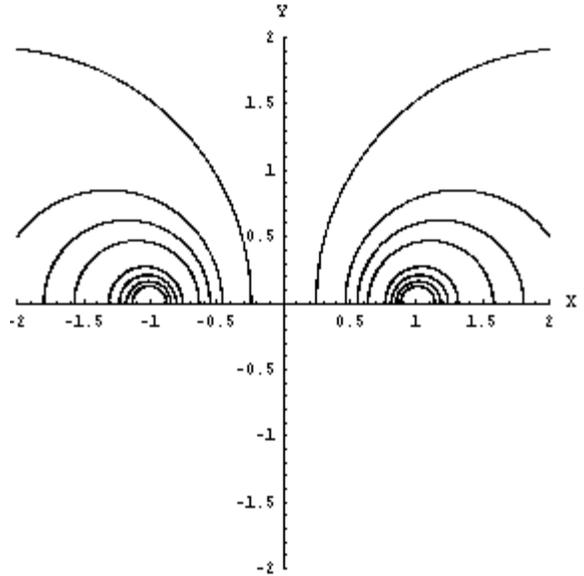

**Fig.3**. Geodesics of the Bipolar Model of the hyperbolic plane

$$r_x(\phi) = h \left( \cos \phi \coth \eta - \sqrt{\cos^2 \phi \coth^2 \eta - 1} \right) \qquad (8)$$

The Cartesian equation of these "x-circles" is given by:

$$(x - h\coth\eta)^2 + y^2 = h^2\text{csch}^2\eta \qquad (9)$$

We demonstrate below that this family of circles has, in fact, *two centers*: a Euclidean center at $x = h\coth\eta$ and a hyperbolic center at the pole $x = h = 1$.

Under the transformation $(\eta, \psi) \to (\psi, \eta)$, retaining the definition of the bipolar coordinates in Eq. (5), metric (7) becomes:

$$ds^2 = \frac{d\eta^2 + d\psi^2}{(\sin\eta)^2} \qquad (10)$$

The geodesics of the "complimentary" metric space (10) yield circles centered on the y-axis. Their polar representation is given by:

$$r_y(\phi) = h\left(\sin\phi \cot\psi - \sqrt{\sin^2\phi \cot^2\psi + 1}\right) \qquad (11)$$

These "y-circle" geodesics pass through both poles and are shown in Fig. 4.

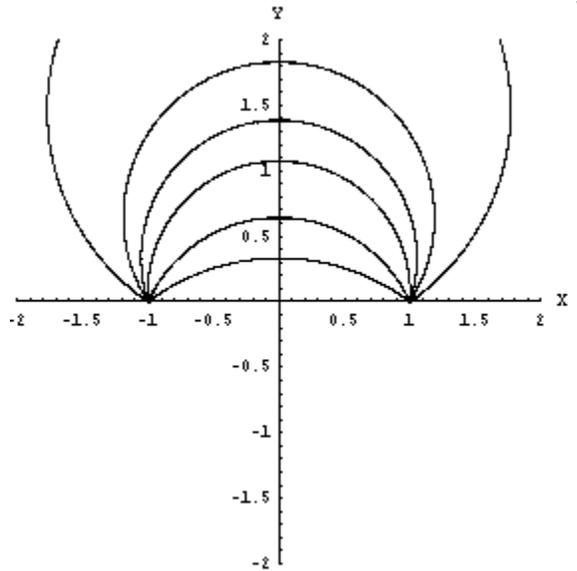

**Fig.4**. Geodesics of the complimentary bipolar hyperbolic metric, Eq. (10).

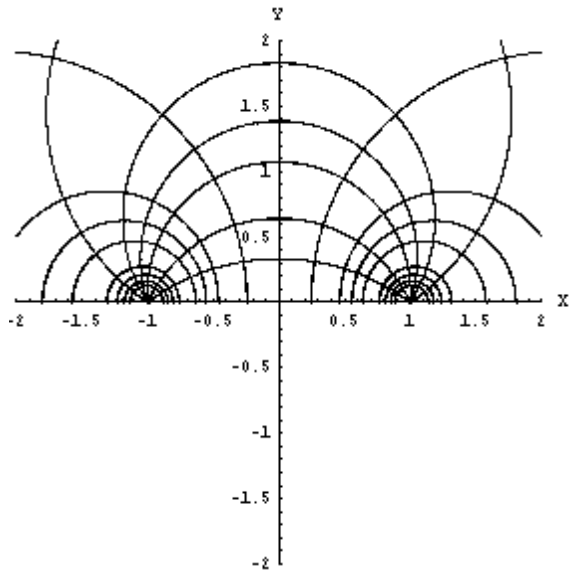

**Fig.5**. Geodesics of Fig.3 and Fig. 4 overlaid.

These circles are orthogonal to the circles (9) and are given in Cartesian coordinates by:

$$(y - h\cot\psi)^2 + x^2 = h^2\csc^2\psi \qquad (12)$$

With respect to the metric space (10), the geodesics of its complimentary space (7) are simply Euclidean circles and therefore, by a fundamental theorem of hyperbolic geometry[8], are also hyperbolic circles about their hyperbolic center. Overlaying the geodesics of the two complimentary spaces in the same Cartesian system yields Fig. 5.

One must, of course, make a choice as to which metric, (7) or (10), will govern the space of the full set of orthogonal curves which coincide with the natural orthogonal bipolar coordinate Euclidean space. For present purposes, we choose metric (10). Then the y-circles are geodesics while the x-circles are Euclidean circles described by Eq. (9) as illustrated in Fig. 6. This graph then becomes, in essence, a demonstration of the "standard construction[9]" required in the well-known proof in hyperbolic geometry that Euclidean circles in the hyperbolic plane are also hyperbolic circles . To this end, it can be shown with respect to metric (10) that the "curved" radii from the hyperbolic center at the pole to a given circle are constant and are given by:

$$\mathbb{R} = \frac{\lambda}{2} \log\left(\frac{\lambda + \rho}{\lambda - \rho}\right) \qquad (13)$$

where $\rho$ is the Euclidean radius of that circle found from Eq.(9). This is shown in Fig. 6

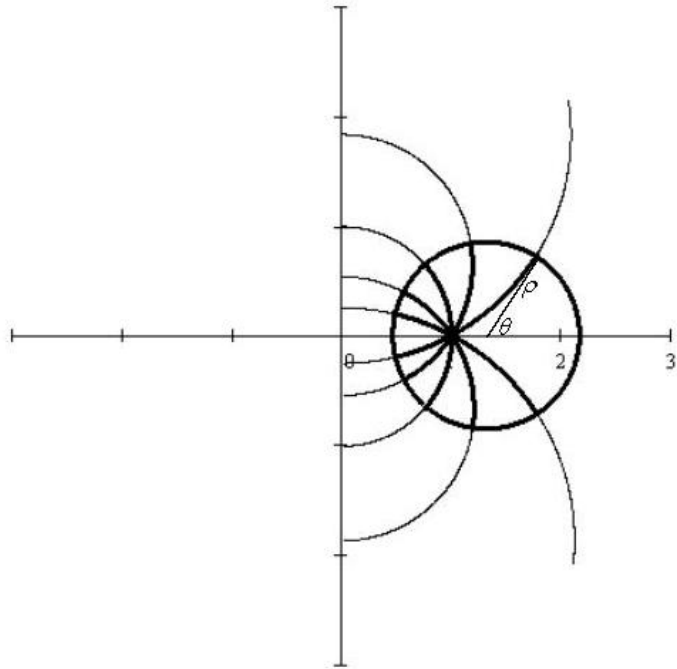

**Fig.6**. Constant radii y-circle geodesics of metric (10) from a pole at $x = 1$ to an x-circle intercept. The Euclidean radius of the circle and angle from the Euclidean center are also shown.

## IV. RELATION TO ROBERTSON-WALKER METRICS

The space inside the Euclidean circle of Fig.6, when referenced to its Euclidean center, can be covered by the the Poincare` disk model[9]. The Poincare` disk is the RW space for curvature $k < 0$. It's metric with respect to the Euclidean center is

$$ds^2 = \frac{d\rho^2 + \rho^2 d\theta^2}{\left(1 + \frac{k\rho^2}{4}\right)^2}, \quad (14)$$

where $\rho$ and $\theta$ are the Euclidean radius and angle shown in Fig. 6. The physical "proper length" of the radius for $\theta = constant$ (a geodesic) is, from (14),

$$s = \frac{\lambda}{2} Ln\left(\frac{\lambda + \rho}{\lambda - \rho}\right), \quad (15)$$

where $\lambda^2 \equiv -4/k$, and is identical to (13). Thus we have shown for a negatively curved RW space that the "proper", or hyperbolic, radius measured from the Euclidean center of the circle is identical to the hyperbolic length of the curved radii measured from its hyperbolic center. This demonstrates the dual-center aspect of orbits in negatively curved RW spaces.

One more useful isometry regarding the curved radii (13) can be described.
We shall use relation (13), which is the curved radius emanating from the hyperbolic center, to transform (14) to a metric with respect to the hyperbolic center. Equation (13) can also be written:

$$\mathbb{R} = \frac{\lambda}{2} log\left(\frac{\lambda + \rho}{\lambda - \rho}\right) = \lambda \tanh^{-1} \frac{\rho}{\lambda} \quad (16)$$

Solving for $\rho$:

$$\rho = \lambda \tanh\left(\frac{\mathbb{R}}{\lambda}\right) \quad (17)$$

Inserting this into (14) yields:

$$ds^2 = d\mathbb{R}^2 + \lambda^2 \sinh^2\left(\frac{\mathbb{R}}{\lambda}\right) d\phi^2 \quad (18)$$

Defining $\chi \equiv \frac{\mathbb{R}}{\lambda}$ and simplifying (18) yields:

$$ds^2 = \lambda^2 \left(d\chi^2 + \sinh^2 \chi \, d\phi^2\right) \quad (19)$$

This is the familiar 2D hyperbolic form of the spatial RW metric. However, we now understand the meaning of "$\chi$". It is simply the curved hyperbolic radius of Fig. 6 emanating from the hyperbolic center. For positively curved RW spaces, there are still two centers, a Euclidean center and an elliptic center which is now imaginary. Also, it can be shown that the distance between the hyperbolic center and the Euclidean center is given by:

$$\varepsilon = \lambda\left(1 - \text{sech}\,\chi\right) \qquad (20)$$

The bipolar model has served to link several isometries of hyperbolic space depending on a judicious choice of origin for measurement: the half-plane model with respect to $(x, y) = (0, 0)$; the bipolar model with respect to $(x, y) = (h, 0)$; and the disk model (RW space) with respect to $(x, y) = (h \coth \eta, 0)$. Each model has its own metric. Only their radii are scaled since these models are conformal.

## V. CONCLUSIONS

Negatively curved spaces must eventually play a role in Cosmology – if for no other reason than void regions in the currently accepted large-scale structure of space must be negatively curved RW space due to the absence of matter. Hyperbolic space is strange and there are a number of models describing various aspects of it. In this work, we have created a new model, which we term the "bipolar model", closely related to a relatively new existing model, the "band model". We have shown that the bipolar model is intimately related to negatively curved Robertson-Walker space and clearly serves to demonstrate an unusual aspect of such spaces – namely that the circular geodesics of such spaces have two real centers, a hyperbolic center as well as a Euclidean center. *These are not the Euclidean center or poles of the bipolar coordinate system but rather refer to two distinct centers for circular orbits of particles in hyperbolic systems, as shown in Fig. 6.* In two dimensions that property is the direct result of the merger of Euclidean plane coordinates and hyperbolic metrics describing the geometry of the "hyperbolic plane". The above work is easily extended to three dimensional space and RW 4-space. Considering the physics of elliptic RW space is so well confirmed in the ΛCDM model with respect to Euclidean coordinates from a Euclidean center, it is likely that the hyperbolic center plays a physical role in regions of hyperbolic space where coordinates are no longer frozen. However that issue is beyond the scope of this paper.

[1] Morse and Feschbach, *Methods of Mathematical Physics, Part II*, (McGraw-Hill, 1953)
[2] Ben M. Leith, S.C. Cindy Ng, and David L. Wiltshire, ApJ **672**, L91 (2008)
[3] H. S. M. Coxeter, *Non-Euclidean Geometry*, Mathematical Association of America (6th Ed., 1998)
[4] L. Randall and R. Sundstrum, Phys. Rev. Lett. **83**, 4690 (1999)
[5] Phillip D. Mannheim, *Brane-Localized Gravity*, (World Scientific Publ., 2005)
[6] H. Ringermacher and L. Mead, J. Math. Phys **46**, 102501 (2005).
[7] John H. Hubbard, *Teichmüller Theory*, (Matrix Editions, Ithaca, NY, 2006)
[8] Saul Stahl, *The Poincaré Half-Plane*, (Jones and Bartlett Publ., 1993)

[9] James W. Cannon, William J. Floyd, Richard Kenyon and Walter R. Parry, Hyperbolic Geometry, *Flavors of Geometry*, (MRSI Publications, Vol. 31, 1997)